\documentclass[12pt]{iopart}

\usepackage{iopams}

\input epsf

\usepackage{graphicx}

\newcommand{\beq}{\begin{equation}}
\newcommand{\eeq}{\end{equation}}
\newcommand{\beqs}{\begin{eqnarray}}
\newcommand{\eeqs}{\end{eqnarray}}
\eqnobysec

\begin{document}

\title[Chromatic Polynomials of Planar Triangulations, the Tutte Upper Bound,
and Chromatic Zeros]{Chromatic Polynomials of Planar Triangulations, the Tutte
Upper Bound, and Chromatic Zeros}

\author{Robert Shrock and Yan Xu}

\address{C. N. Yang Institute for Theoretical Physics \\
Stony Brook University \\ Stony Brook, NY 11794}


\begin{abstract}

Tutte proved that if $G_{pt}$ is a planar triangulation and $P(G_{pt},q)$ is
its chromatic polynomial, then $|P(G_{pt},\tau+1)| \le (\tau-1)^{n-5}$, where
$\tau=(1+\sqrt{5} \, )/2$ and $n$ is the number of vertices in $G_{pt}$. Here
we study the ratio $r(G_{pt})=|P(G_{pt},\tau+1)|/(\tau-1)^{n-5}$ for a variety
of planar triangulations. We construct infinite recursive families of planar
triangulations $G_{pt,m}$ depending on a parameter $m$ linearly related to $n$
and show that if $P(G_{pt,m},q)$ only involves a single power of a polynomial,
then $r(G_{pt,m})$ approaches zero exponentially fast as $n \to \infty$. We
also construct infinite recursive families for which $P(G_{pt,m},q)$ is a sum
of powers of certain functions and show that for these, $r(G_{pt,m})$ may
approach a finite nonzero constant as $n \to \infty$.  The connection between
the Tutte upper bound and the observed chromatic zero(s) near to $\tau+1$ is
investigated.  We report the first known graph for which the zero(s) closest to
$\tau+1$ is not real, but instead is a complex-conjugate pair.  Finally, we
discuss connections with nonzero ground-state entropy of the Potts
antiferromagnet on these families of graphs.

\end{abstract}



\maketitle

\newpage
\pagestyle{plain}
\pagenumbering{arabic}
\section{Introduction}
\label{intro}

The chromatic polynomial is a function of interest in both graph theory and
statistical physics. Let $G=(V,E)$ be a graph with vertex and edge sets $V$ and
$E$, and denote the number of vertices and edges as $n=n(G)=|V|$ and
$e(G)=|E|$, respectively. The chromatic polynomial $P(G,q)$ counts the number
of ways of assigning $q$ colors to the vertices of $G$, subject to the
condition that no two adjacent vertices have the same color
\cite{biggsbook,dong}.  Such an assignment is called a proper $q$-coloring of
$G$.  The minimum number of colors for a proper $q$-coloring of a graph $G$ is
the chromatic number of $G$, denoted $\chi(G)$.  The connection to statistical
physics arises from the fact that $P(G,q)=Z_{PAF}(G,q,0)$, where
$Z_{PAF}(G,q,T)$ denotes the partition function of the Potts antiferromagnet
(PAF) on the graph $G$ at temperature $T$ \cite{wurev}.  A further important
aspect of this connection involves nonzero ground-state entropy, $S_0 >
0$. This is equivalent to a ground state degeneracy per site $W > 1$, since
$S_0 = k_B \ln W$.  The Potts antiferromagnet exhibits nonzero ground-state
(i.e., zero-temperature) entropy for sufficiently large $q$ on a given graph
and serves as a useful model for the study of this phenomenon.  The chromatic
polynomial of a graph $G$ may be computed via the deletion-contraction relation
$P(G,q)=P(G-e,q)-P(G/e,q)$, where $G-e$ denotes $G$ with an edge $e$ deleted
and $G/e$ denotes the graph obtained by deleting $e$ and identifying the
vertices that it connected.  From this follows the general formula $P(G,q) =
\sum_{G' \subseteq G} (-1)^{e(G')} \ q^{k(G')}$, where $G'=(V,E')$ with $E'
\subseteq E$ and $k(G')$ denotes the number of connected components of $G'$.

While the chromatic polynomial $P(G,q)$ enumerates proper $q$-colorings for
positive integer $q$, it is well-defined for any (real or complex) $q$.  For an
arbitrary $G$ and arbitrary $q$, the calculation of $P(G,q)$ is, in general, an
exponentially hard problem.  Hence, bounds on $P(G,q)$ are of great value.  An
interesting upper bound on an evaluation of chromatic polynomials was proved by
Tutte.  Let us consider a graph $G_{pt}$ that is a planar triangulation ($pt$),
defined as a graph that can be drawn in a plane without any intersecting edges
and has the property that all of its faces are triangles. This is 
necessarily a connected graph. Tutte proved an upper bound on the
absolute value of the chromatic polynomial of a planar triangulation,
$P(G_{pt},q)$ evaluated at a certain value of $q$, viz.,
\beq
q =\tau+1= \frac{3+\sqrt{5}}{2} = 2.6180339887...
\label{tp1}
\eeq
where $\tau=(1+\sqrt{5} \, )/2$ is the ``golden ratio'', satisfying
$\tau+1=\tau^2$, or equivalently, $\tau^{-1} = \tau-1$. Tutte's upper bound is
\cite{tutte70} (see also \cite{tutte69},\cite{tutte84}),
\beq
|P(G_{pt}, \tau+1)| \le U(n(G_{pt})) \ , 
\label{tub}
\eeq
where
\beq 
U(n) = \tau^{5-n} = (\tau-1)^{n-5} \ .
\label{u}
\eeq
Since $\tau-1 < 1$, the upper bound $U(n)$ decreases exponentially rapidly as a
function of $n$.  For a planar triangulation graph $G_{pt}$, we define the
ratio of the evaluation of its chromatic polynomial at $q=\tau+1$ to the Tutte
upper bound as
\beq
r(G_{pt}) \equiv \frac{|P(G_{pt},\tau+1)|}{U(n(G_{pt}))} \ . 
\label{rpg}
\eeq

In this paper we study this ratio $r(G_{pt})$ for various planar triangulation
graphs.  For this purpose, we shall construct infinite recursive families of
planar triangulations $G_{pt,m}$ depending on a parameter $m$ linearly related
to $n$ and show that if $P(G_{pt,m},q)$ only involves a single power of a
polynomial, $[\lambda_{G_{pt}}(q)]^m$, then $r(G_{pt,m})$ approaches zero
exponentially fast as $m \to \infty$. We also construct infinite recursive
families for which $P(G_{pt,m},q)$ is a sum of powers of certain functions and
show that for these, $r(G_{pt,m})$ may approach a finite nonzero constant as $m
\to \infty$.  The Tutte upper bound is sharp, since it is saturated by the
simplest planar triangulation, namely the triangle, $K_3$ \footnote{The
complete graph $K_n$ is defined as the graph with $n$ vertices such that every
vertex is connected by an edge to every other vertex.  Its chromatic polynomial
is $P(K_n,q)=\prod_{s=0}^{n-1}(q-s)$, and its chromatic number is
$\chi(K_n)=n$.}. For this graph,
\beq
P(K_3,\tau+1)=\tau+1  \ ,  \quad \Rightarrow \ \ r(K_3)=1 \ .  
\label{pk3rk3}
\eeq
However, for planar triangulations with higher $n$, the upper bound is realized
as a strict inequality.  For example, for $K_4$,
\beq
P(K_4,\tau+1)= -1, \quad \Rightarrow \ \ r(K_4) = \tau-1 = 0.61803..
\label{pk4rk4}
\eeq

In connection with the upper bound (\ref{tub}), Tutte reported his empirical
observation that chromatic polynomials of planar triangulations typically have
a real zero quite close to $q=\tau+1$ and remarked that this property could be
related to the fact that they obey his bound \cite{tutte70}-\cite{tutte84}.
Although several decades have passed since Tutte's work on this topic, the
nature of this relation between the zero(s) of a chromatic polynomial
$P(G_{pt},q)$ near to $q=\tau+1$ and the upper bound (\ref{tub}) remains to be
understood.  We shall elucidate this connection and generalize the
investigation to the study of complex zeros of $P(G_{pt},q)$.  In particular,
we report the first known case, to our knowledge, of a planar triangulation for
which the zero(s) closest to $\tau+1$ is not a real one, but instead, is a
complex-conjugate pair.

Before proceeding, we include some additional notation and remarks.  Since a
triangulation $G_t$ (whether planar or not) contains at least one triangle,
$P(G_t,q)$ always contains the factor $P(K_3,q) = q(q-1)(q-2)$.  Tutte
emphasized the relation of Eq. (\ref{tub}) to the golden ratio, $\tau$.  It is
also of interest to note a relation with certain roots of unity, namely
$z_r=e^{\pi i/r}$.  One defines $q_r = (z_r+z_r^*)^2 = 4\cos^2(\pi/r)$, which
are often called Tutte-Beraha numbers.  The connection with the golden ratio
$\tau$ is that $q_5 = \tau+1$. 

\section{Recursive Families of Planar Triangulations}
\label{recursive}

We have studied various families of planar triangulations that can be
constructed in a recursive manner.  We denote the $m$'th member of such a
family as $G_{pt,m}$ and the number of vertices as
\beq
n(G_{pt,m}) = \alpha \, m + \beta \ , 
\label{nmrel}
\eeq
where $\alpha$ and $\beta$ depend on the type of family. Since $U(n) \to 0$ as
$n \to \infty$ and since $m$ is proportional to $n$, it follows, first, that
for these recursive families of planar triangulations,
\beq
\lim_{m \to \infty} P(G_{pt,m},\tau+1) = 0 \ . 
\label{ptp1limit}
\eeq
Second, given the upper bound (\ref{tub}) and the fact that $U(n)$ approaches
zero exponentially fast as $n \to \infty$, it follows that 
\beq
P(G_{pt,m},\tau+1) \quad {\rm approaches \ \ zero \ \ exponentially \ \ fast
  \ \ as} \ \ m \to \infty . 
\label{ptp1approach}
\eeq

There are several ways of constructing recursive families of planar
triangulations.  We have used the following method to construct families for
which the chromatic polynomial involves a single power (of a polynomial in
$q$). Start with a basic graph $G_{pt,1}$, drawn in the usual explicitly planar
manner.  The outer edges of this graph clearly form a triangle, $K_3$.  Next
pick an interior triangle in $G_{pt,1}$ and place a copy of $G_{pt,1}$ in this
triangle so that the intersection of the resultant graph with the original
$G_{pt,1}$ is the triangle chosen.  Denote this as $G_{pt,2}$. Continuing in
this manner, one constructs $G_{pt,m}$ with $m \ge 3$.  The chromatic
polynomial $P(G_{pt,2},q)$ is calculated from $P(G_{pt,1},q)$ by using the
complete graph intersection theorem.  This theorem states that if for two
graphs $G$ and $H$ (which are not necessarily planar or triangulations), the 
intersection $G \cap H = K_p$ for some $p$, then
\beq
P(G \cup H,q) = \frac{P(G,q)P(H,q)}{P(K_p,q)} = 
\frac{P(G,q)P(H,q)}{\prod_{s=0}^{p-1}(q-s)} \ . 
\label{cgit}
\eeq
For our planar triangulations, we use the $p=3$ special case of this theorem
and take advantage of the property that $H$ is a copy of $G$.  Consequently,
for planar triangulations formed in this recursive manner, the chromatic
polynomial has the form
\beq
P(G_{pt,m},q) = P(K_3,q) \, [\lambda_{G_{pt}}(q)]^m \ , 
\label{pgform1}
\eeq
The chromatic number of $G_{pt}$ may be 3 or 4; if $\chi(G_{pt})=4$, then
$\lambda_{G_{pt}}(q)$ contains the factor $q-3$.  In the case of a planar
triangulation which is a strip of the triangular lattice of length $m$ vertices
with cylindrical boundary conditions, to be discussed below, an alternate and
equivalent way to construct the $(m+1)$'th member of the family is simply to
add a layer of vertices to the strip at one end.

We have also devised a method to obtain recursive families of planar
triangulations with the property that the chromatic polynomial is a sum of more
than one power of a function of $q$, which we generically write as 
\beq
P(G_{pt,m},q) = \sum_{j=1}^{j_{max}} c_{G_{pt},j}(q) [\lambda_{G_{pt},j}(q)]^m \ , 
\label{pgform}
\eeq
where $j_{max} \ge 2$, the $c_{G_{pt},j}(q)$ are certain coefficients, and we use
the label $G_{pt}$ to refer to the general family of planar triangulations
$G_{pt,m}$.  We will describe this method in detail below.  Parenthetically, we
recall that the form (\ref{pgform}) is a general one for recursive families of
graphs, whether or not they are planar triangulations \cite{bds}, \cite{w}.
For (\ref{pgform}) evaluated at a given value $q=q_0$, as $m \to \infty$, and
hence $n \to \infty$, the behavior of $P(G_{pt,m},q)$ is controlled by which
$\lambda_{G_{pt},j}(q)$ is dominant at $q=q_0$, i.e., which of these has the
largest magnitude $|\lambda_{G_{pt},j}(q_0)|$.  For our purposes, the $q_0$ of
interest is $\tau+1$. We denote the $\lambda_{G_{pt},j}(q)$ that is dominant at
$q=\tau+1$ as $\lambda_{G_{pt},dom}(\tau+1)$. Clearly, if $P(G_{pt,m},q)$
involves only a single power, as in (\ref{pgform1}), then
$\lambda_{G_{pt},dom}(\tau+1)=\lambda_{G_{pt}}(\tau+1)$.

Thus, in both the cases of Eq. (\ref{pgform1}) and (\ref{pgform}), a single
power $[\lambda_{G_{pt},j}(q)]^m$ dominates the sum as $m \to \infty$.  The
upper bound $U(n(G_{pt,m}))=(\tau-1)^{n-5}$ also has the form of a power, and
therefore it is natural to define a (real, non-negative) constant $a_{G_{pt}}$ as
\beq
a_{G_{pt}} = \lim_{n \to \infty} [ r(G_{pt,m}) ]^{1/n} = 
\frac{|\lambda_{G_{pt},dom}(\tau+1)|^{1/\alpha}}{\tau-1} \ . 
\label{a}
\eeq
To study the asymptotic behavior of $r(G_{pt,m})$ as $m$ and hence $n$ go to
infinity, and to determine the resultant constant $a_{G_{pt}}$, we consider the
two different cases in turn, namely those in Eqs. (\ref{pgform1}) and
(\ref{pgform}). 

We begin with the first case, viz., that of Eq. (\ref{pgform1}). Because
$r(G_{pt,m})$ must satisfy the upper bound (\ref{tub}) for any $m$, and because
this is satisfied as a strict inequality for all planar triangulations other
than the simplest, $K_3$, it follows that
\beq
a_{G_{pt}} < 1 \ . 
\label{aineq}
\eeq
Equivalently, 
\beq
|\lambda_{G_{pt}}(\tau+1)|^{1/\alpha} < \tau-1 \ . 
\label{fineq1}
\eeq
For all of the $G_{pt,m}$ studied here, $\alpha \ge 1$, so (\ref{fineq1})
implies
\beq
|\lambda_{G_{pt}}(\tau+1)| < \tau-1 \ . 
\label{fineq2}
\eeq
From this, we derive two important results, namely that for these 
recursive families of planar triangulation graphs $G_{pt,m}$ where 
$P(G_{pt,m},q)$ has the form (\ref{pgform1}) involving only a single power 
$[\lambda_{G_{pt}}(q)]^m$, first, 
\beq
\lim_{n \to \infty} r(G_{pt,m}) = 0 \ , 
\label{rinfform1}
\eeq
and second, 
\beq
r(G_{pt,m}) \ \ {\rm decreases \ exponentially \ fast \ as \ a \ 
function \ of} \ m \ {\rm and} \ n \ . 
\label{rexponentialdecay1}
\eeq

In contrast, for recursive families of planar triangulation graphs $G_{pt,m}$
where $P(G_{pt,m},q)$ has the form (\ref{pgform}) involving two or more powers
$[\lambda_{G_{pt},j}(q)]^m$, it is possible for $\lim_{n \to \infty}
r(G_{pt,m})$ to be a nonzero constant (which necessarily lies in the interval
$(0,1)$), so that $a_{G_{pt}}=1$.  This type of behavior means that if one sets
$q = \tau+1$ and takes $n \to \infty$, the $\lambda_{G_{pt},j}(q)$ that enters
in the dominant term in (\ref{pgform}) satisfies
\beq
\lambda_{G_{pt},dom}(\tau+1)=\tau-1 \quad {\rm for} \ \ a_{G_{pt}}=1 \ .
\label{lamdom}
\eeq

\section{The Family $R_m = P_m + P_2$}
\label{Rm}

\begin{figure}
  \begin{center}
    \includegraphics[height=6cm]{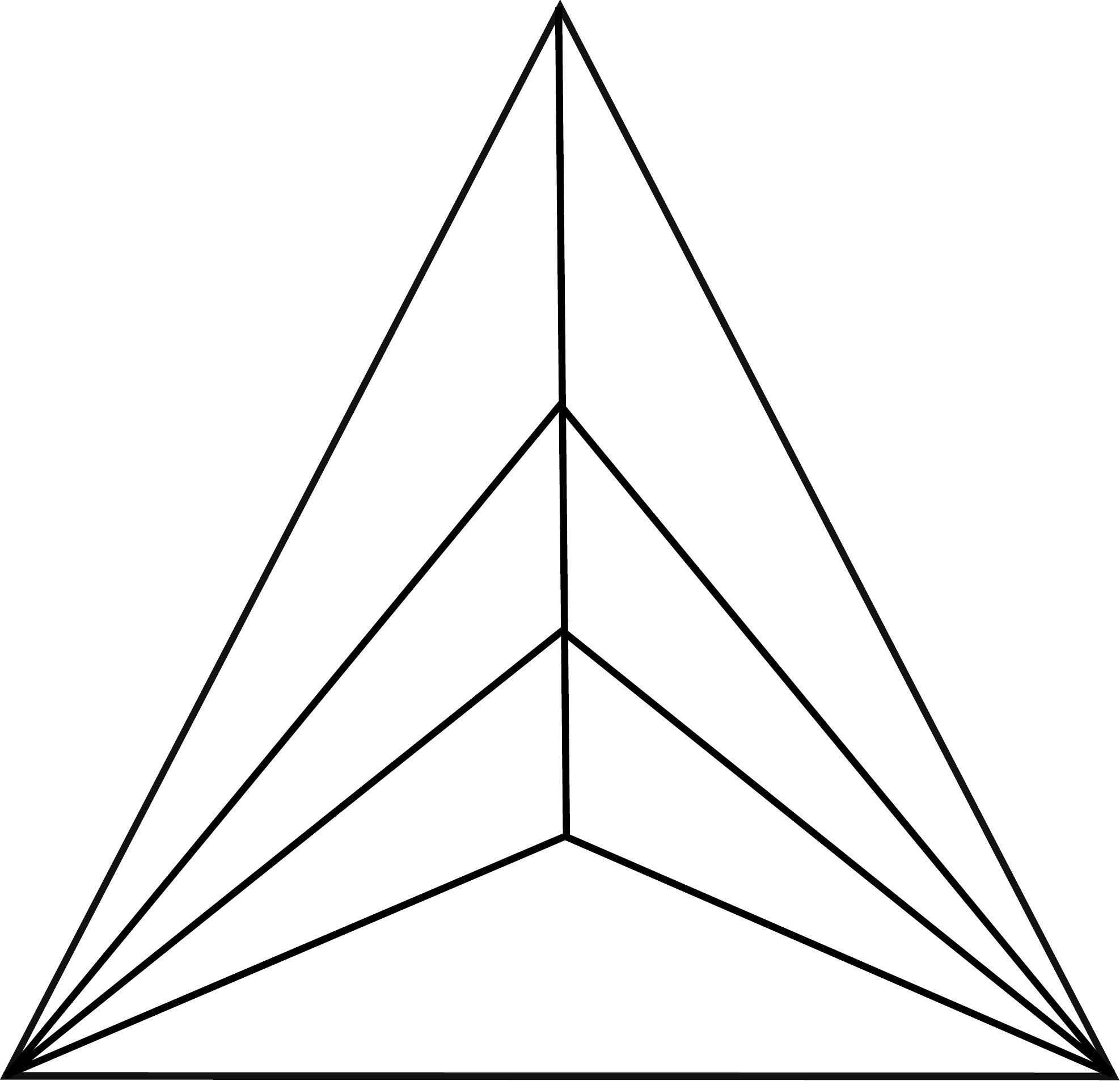}
  \end{center}
\caption{Graph $R_4$.}
\label{R4}
\end{figure}

In this section we illustrate the general results derived in the previous
section for recursive families of planar triangulations whose chromatic
polynomials have the form (\ref{pgform1}).  For this purpose, we consider a
family whose $m$'th member, denoted $R_m$, is the join\footnote{ The join $G+H$
of two graphs $G=(V_G,E_G)$ and $H=(V_H,E_H)$ is defined as the graph with
vertex set $V_{G+H}=V_G \cup V_H$ and edge set $E_{G+H}$ comprised of the union
of $E_G \cup E_H$ with the set of edges obtained by connecting each vertex of
$G$ with each vertex of $H$.}  of the path graph $P_m$ with $P_2=K_2$,
\beq
R_m = P_m + P_2 \ .
\label{trm}
\eeq
Thus, $R_1=K_3$, $R_2=K_4$, etc. The graph $R_4$ is shown in Fig. \ref{R4}. 
We have $n(R_m) = m+2$, so $\alpha=1$ and $\beta=2$ for this family.  An
elementary calculation yields $P(R_m,q) = q(q-1)(q-2)[\lambda_R(q)]^{m-1}$,
where $\lambda_R(q) = q-3$.  Evaluating $P(R_m,q)$ at $q=\tau+1$, we obtain
\beq
P(R_m,\tau+1) = (\tau+1)(\tau-2)^{m-1} = \Big ( \frac{3+\sqrt{5}}{2} \ \Big ) 
          \Big ( \frac{-3+\sqrt{5}}{2} \ \Big )^{m-1}
\label{ptrmev}
\eeq
The ratio of $|P(R_m,\tau+1)|$ to the Tutte upper bound is 
\beqs
r(R_m) & = & (\tau-1)^{m-1} = \Big ( \frac{-1+\sqrt{5}}{2} \Big )^{m-1} 
\label{rtrm}
\eeqs
so that 
\beq
a_R = \frac{-1+\sqrt{5}}{2} = 0.61803..
\label{arm}
\eeq
for this family.

\section{The Cylindrical Strip of the Triangular Lattice with $L_y=3$}
\label{TCm} 

Another illustration of recursive families of planar triangulations whose
chromatic polynomials have the form (\ref{pgform1}) is provided by the family
of cylindrical strips of the triangular ($tri$) lattice. Consider a strip in
this family of variable length $L_x=m$ vertices in the $x$ (longitudinal)
direction and $L_y$ vertices in the $y$ (transverse) direction. Denote the
boundary conditions in the $x$ and $y$ directions as $(BC_x, \ BC_y)$, where
free and periodic boundary conditions are abbreviated as $FBC$ and $PBC$.  We
take the boundary conditions to be $(FBC_x, \ PBC_y)$, denoted as cylindrical.
The cylindrical strip of the triangular lattice with $L_y=3$, i.e., transverse
cross sections consisting of triangles, and arbitrary length $L_x=m$, denoted
$TC_m$, is a planar triangulation, with $n(TC_m)=3m$ vertices.  In this family,
$TC_1$ is a degenerate case of a triangle, $K_3$, while $TC_2$ is the graph of
the octahedron. $TC_m$ may also be constructed in the recursive manner
described in Sect. \ref{recursive} by starting with $TC_2$, choosing an
interior triangle, and placing a copy of $TC_2$ onto this triangle to get
$TC_3$, and so forth for $m \ge 4$.  Thus, $TC_m$ may also be considered as a
recursively iterated octahedron graph. An elementary calculation yields
$P(TC_m,q) = q(q-1)(q-2) \, [\lambda_{TC}(q)]^{m-1}$, where
$\lambda_{TC}(q)=q^3-9q^2+29q-32$. Evaluating $P(TC_m,q)$ at $q=\tau+1$ gives
\beq
P(TC_m,\tau+1)= \Big ( \frac{3+\sqrt{5}}{2} \ \Big )(-11+5\sqrt{5} \ )^{m-1} 
\ . 
\label{ptcmev}
\eeq
Hence, 
\beq
r(TC_m) = \frac{(3+\sqrt{5} \, )}{4} \ (3-\sqrt{5} \, )^m \ , 
\label{rtcm}
\eeq
so that 
\beq
a_{TC} = (3-\sqrt{5} \, )^{1/3} = 0.91415...
\label{atc}
\eeq
$P(TC_m,q)$ has a real zero near to $\tau+1$, at $q \simeq 2.546602$.

\section{Iterated Icosahedron Graphs}
\label{Im}

\begin{figure}
  \begin{center}
    \includegraphics[height=6cm]{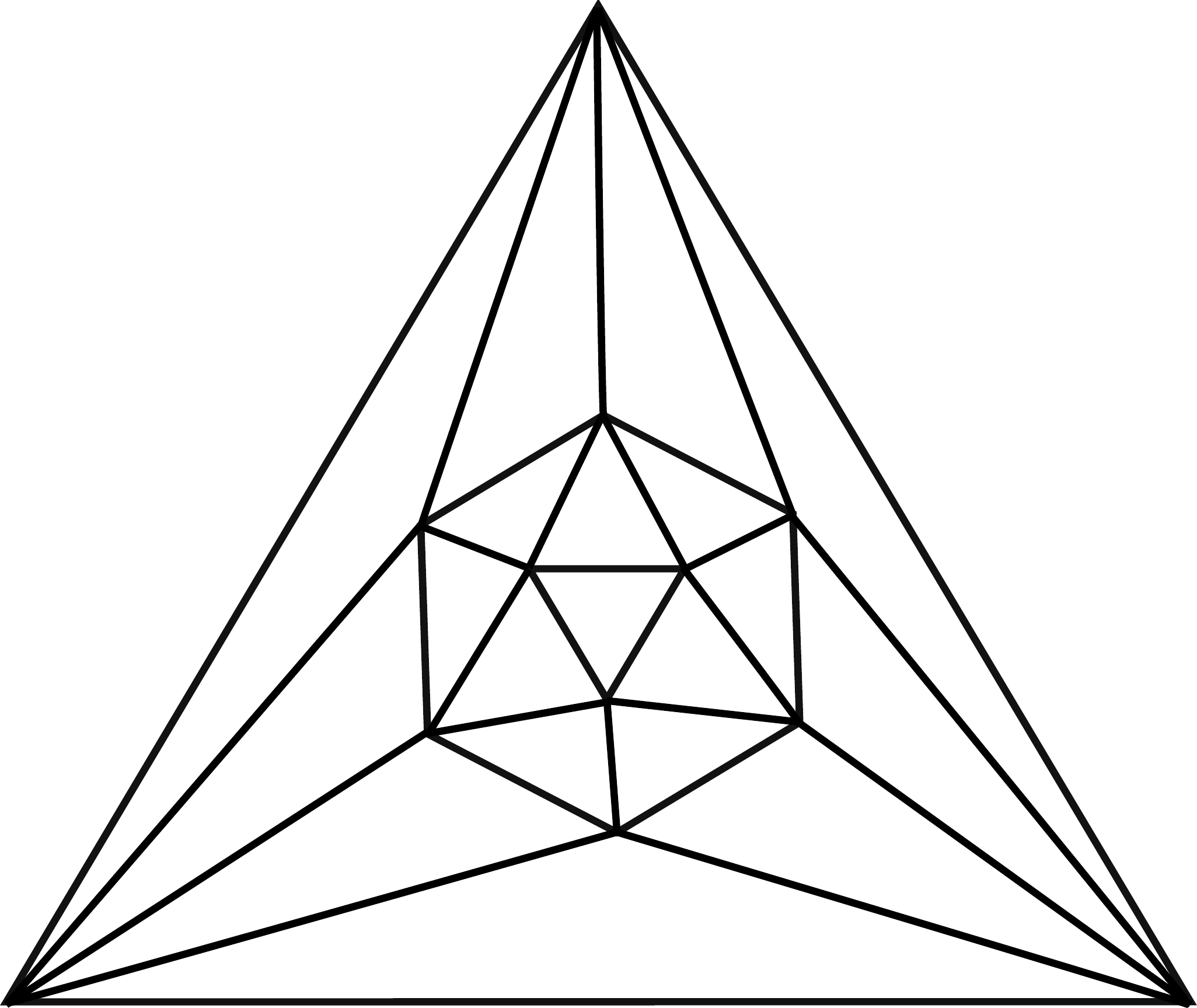}
  \end{center}
\caption{Icosahedron graph, $I_1$.}
\label{icosahedron}
\end{figure}

A more complicated family of planar triangulations yielding chromatic
polynomials of the form (\ref{pgform1}) is provided by recursive iterates of
the icosahedron graph $I_1$, shown in Fig. \ref{icosahedron}.  We construct
these recursive iterates $I_m$ with $m \ge 2$ by the procedure given in
Sect. \ref{recursive}. The graph $I_m$ has $n(I_m)=3+9m$ vertices.
The chromatic polynomial is
\beq
P(I_m,q) = q(q-1)(q-2) \, [f_I(q)]^m
\label{p_35}
\eeq
where
\beqs 
f_I(q) & = & (q-3)(q^8-24q^7+260q^6-1670q^5+6999q^4-19698q^3 \cr\cr
          & + & 36408q^2 -40240q+20170) \ . 
\label{fi}
\eeqs
We calculate 
\beq
P(I_m,\tau+1)= \Big ( \frac{3+\sqrt{5}}{2} \ \Big ) \, 
\Big ( \frac{-23955+10713\sqrt{5}}{2} \ \Big )^m \ . 
\label{pimev}
\eeq
Hence, 
\beq
r(I_m) = \frac{(3+\sqrt{5} \, )}{4} \, \Big ( \frac{-315+141\sqrt{5}}{2} \ 
\Big )^m \ , 
\label{rim}
\eeq
so that 
\beq
a_{I} =\Big ( \frac{-315+141\sqrt{5} \ }{2} \Big )^{1/9} = 0.80552...
\label{aim}
\eeq
$P(I_m,q)$ has a real zero quite close to $\tau+1$, at $q \simeq
2.6181973$ and another real zero near to $q_7$, at $q \simeq 3.222458$.

\section{The Family $B_n$ }

\begin{figure}
  \begin{center}
    \includegraphics[height=6cm]{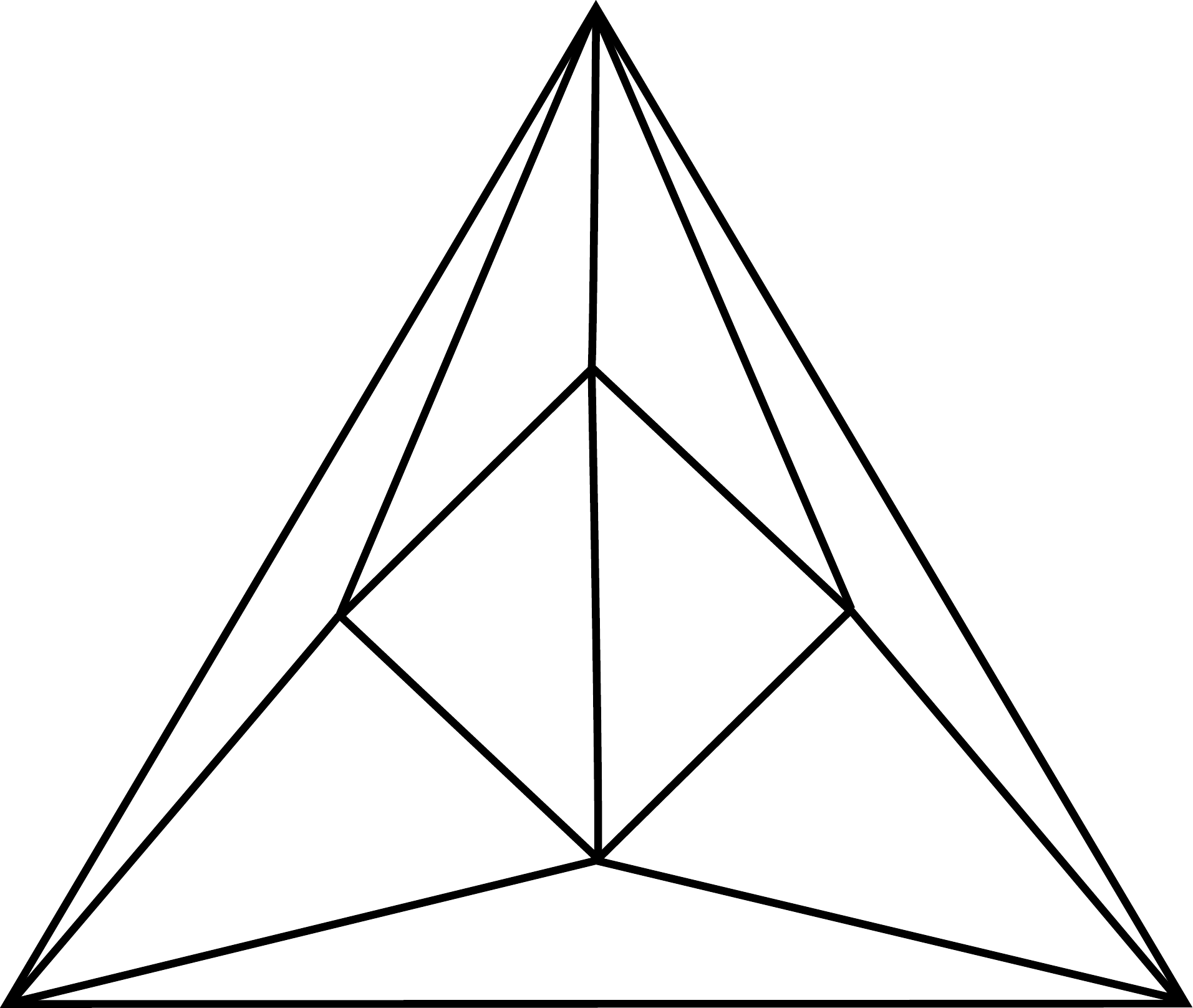}
  \end{center}
\caption{Graph $B_7$.}
\label{B7}
\end{figure}

\begin{figure}
  \begin{center}
    \includegraphics[height=6cm]{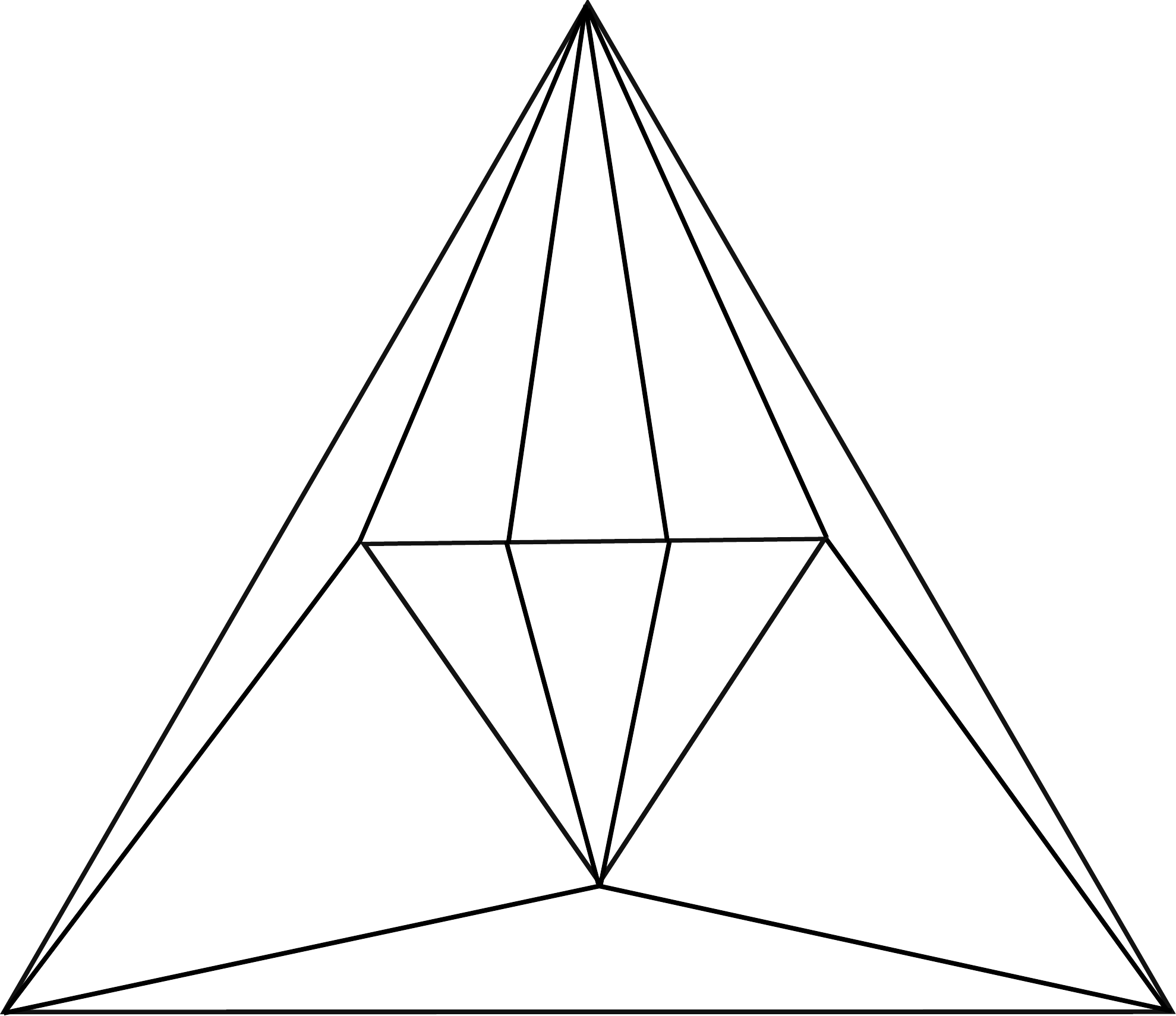}
  \end{center}
\caption{Graph $B_8$.}
\label{B8}
\end{figure}

In this section we consider a recursive family of planar triangulations whose
chromatic polynomials have the form (\ref{pgform}) with $j_{max}=3$. This is
the bipyramid family, which is the join
\beq
B_n = \bar K_2 + C_{n-2}
\eeq
for $n \ge 5$, where $\bar K_p$, the complement of $K_p$, is the graph of $p$
disjoint vertices with no edges.  Clearly, $n(B_n)=n$.  For illustration, the
graphs $B_7$ and $B_8$ are shown in Figs. \ref{B7} and \ref{B8}.
In Figs. \ref{B7} and \ref{B8}, the uppermost and the lower middle vertices of
$B_n$ have degree $n-2$, so that the degrees of these two vertices go to
infinity as $n \to \infty$.  All of the other vertices have degree 4.  The
chromatic polynomial for this graph is of the form (\ref{pgform}) with three
$\lambda$s:
\beq
P(B_n,q) = \sum_{j=1}^3 c_{B,j}(q)[\lambda_{B,j}(q)]^{n-2} \ , 
\label{pbn}
\eeq
where
\beq
c_{B,j}(q) = q, \quad c_{B,2}(q) = q(q-1), \quad c_{B,3}(q) =q(q^2-3q+1)
\label{cjbn}
\eeq
and
\beq
\lambda_{B,1}(q) = q-2, \quad \lambda_{B,2}(q)=q-3, \quad 
\lambda_{B,3}(q)=-1 \ . 
\label{lambdabn}
\eeq
$P(B_n,q)$ contains the factor $P(K_3,q)$ if $n$ is even and $P(K_4,q)$ if $n$
is odd; related to this, $\chi(B_n)=3$ if $n$ is even and $\chi(B_n)=4$ if
$n$ is odd. 

Evaluating $P(B_n,q)$ at $q=\tau+1$, we find 
\beq
P(B_n,\tau+1) = (\tau+1)\Big [ (\tau-1)^{n-2} + \tau(\tau-2)^{n-2} \Big ] \ . 
\label{pbntp1}
\eeq
Note that $c_{B,3}(q)$ vanishes for $q=\tau+1$ (see the related Eq. (2.8) of
\cite{cf}), so that the $j=3$ term does not contribute
for this value of $q$. Hence,
\beq
r(B_n) = (\tau-1) \Big [ 1 + \tau(1-\tau)^{n-2} \Big ]
\label{rbn}
\eeq
Since $|1-\tau| < 1$, the second term, $\tau(1-\tau)^{n-2}$, vanishes as $n \to
\infty$, so 
\beq
\lim_{n \to \infty} r(B_n) = \tau-1 = \frac{-1+\sqrt{5}}{2} = 0.61803..
\label{rbninf}
\eeq
and
\beq
a_B = 1 \ . 
\label{ab}
\eeq
Because $1-\tau$ is negative, the ratios $r(B_n)$ form a sequence such that for
increasing even (odd) $n$, $r(B_n)$ approaches the $n \to \infty$ limit
$\tau-1$ from above (below).  For $n \ge 6$, $P(B_n,q)$ has real zeros that are
close to $q=\tau+1$.  These form a sequence such that for even (odd) $n$ the
nearby zero is slightly less than (greater than) $q=\tau+1$,
respectively. These are listed in Table \ref{bnzerotable} for $n$ from 6 to 20.
\begin{table}[htbp]
\caption{\footnotesize{Location of zero $q_z$ of $P(B_n,q)$ closest to
   $\tau+1=2.6180339887..$ for $n$ from 6 to 20.  Notation $a$e-n means 
$a \times 10^{-n}$.}}
\begin{indented}
\item[]\begin{tabular}{|c|c|c|}
\br
$n$ &  $q_z$     & $q_z-(\tau+1)$  \\
\mr
6   &  2.546602   &  $-0.07143$    \\
7   &  2.677815   &  0.05978       \\
8   &  2.594829   &  $-0.02321$    \\
9   &  2.636118   &  0.01808       \\
10  &  2.609130   &  $-0.8904$e-2  \\
11  &  2.624356   &  0.6322e-2     \\
12  &  2.614541   &  $-3.493$e-3   \\
13  &  2.620356   &  2.322e-3      \\
14  &  2.616673   &  $-1.361$e-3   \\
15  &  2.618905   &  0.8713e-3     \\
16  &  2.617509   &  $-0.5254$e-3  \\
17  &  2.618364   &  3.301e-4      \\
18  &  2.617832   &  $-2.017$e-4   \\
19  &  2.618160   &  1.256e-4      \\
20  &  2.617957   &  $-0.7725$e-4  \\
\br
\end{tabular}
\end{indented}
\label{bnzerotable}
\end{table}

The continuous accumulation set of zeros of $P(B_n,q)$ in the complex $q$ plane
as $n \to \infty$ was studied in \cite{rr,w}.  These form curves defined by the
equality in magnitude of the dominant $\lambda_{B,j}$, in accordance with
general results for recursive functions \cite{bkw}. These curves extend
infinitely far from the origin if such an equality can be satisfied as $|q| \to
\infty$, as was discussed in \cite{w,wa,wa3}. Here, because the equality
$|\lambda_{B,1}|=|\lambda_{B,2}|$, i.e., $|q-2|=|q-3|$, holds for the infinite
line ${\rm Re}(q)=5/2$, part of the boundary ${\cal B}$ extends infinitely far
away from the origin in the $q$ plane, i.e., passes through the origin of the
$1/q$ plane.  The locus ${\cal B}$ separates the $q$ plane into three regions
(see Fig. 2 and Eqs. (4.11)-(4.16) in \cite{w}): (i) region $R_1$, which
includes the semi-infinite interval $q > 3$ on the real axis; (ii) region
$R_2$, which includes the semi-infinite interval $q < 2$ on the real axis; and
(iii) region $R_3$, a region bounded below on the real axis by $q=2$ and above
by $q=3$, and extending upward and downward to the triple points $q = (5 \pm
\sqrt{3} \, i)/2$, where the three regions are contiguous.  Denoting $q_c$ as
the maximal point where ${\cal B}$ intersects the real-$q$ axis, we have
$q_c=3$ here. The boundary between regions $R_1$ and $R_2$ for $|{\rm Im}(q)| >
\sqrt{3} \, /2$ is comprised by semi-infinite segments of the vertical line
${\rm Re}(q)=5/2$.  In region $R_1$, $\lambda_{B,1}$ is dominant. 

In the calculation of the constant $a_{G_{pt}}$ of Eq. (\ref{a}), we note that
one first sets $q=\tau+1$ and then takes $n \to \infty$. In general, for a
special set of values of $q$, denoted $\{q_s\}$, one has the noncommutativity
\cite{w,a}
\beq
\lim_{q \to q_s} \lim_{n \to \infty} [P(G,q)]^{1/n} \ne 
\lim_{n \to \infty} \lim_{q \to q_s} [P(G,q)]^{1/n} \ . 
\label{wnoncom}
\eeq
This noncommutativity is illustrated by the family $B_n$, for which the set of
special values $\{q_s\}$ includes $q=0, \ 1, \ 2, \ 3$, and $\tau+1$.  Thus,
for the evaluation of $a_{B}$, one first sets $q=\tau+1$ (which is a point 
lying in region $R_3$) and then takes $n \to \infty$.  In this case, 
since $c_{B,3}(q)=0$, for $q=\tau+1$, the $(\lambda_{B,3})^{n-2}$
term is multiplied by zero and does not contribute to the chromatic polynomial,
and the resultant dominant term is $\lambda_{B,1}=q-2$, whence the results in 
Eqs. (\ref{rbn}) and (\ref{ab}).  In contrast, if one takes 
$n \to \infty$ first, then the domninant term in region $R_3$ is 
$\lambda_{B,3}=-1$ (and the dominant term in $R_2$ is $q-3$). 

\section{The Family $H_n$}

\begin{figure}
  \begin{center}
    \includegraphics[height=6cm]{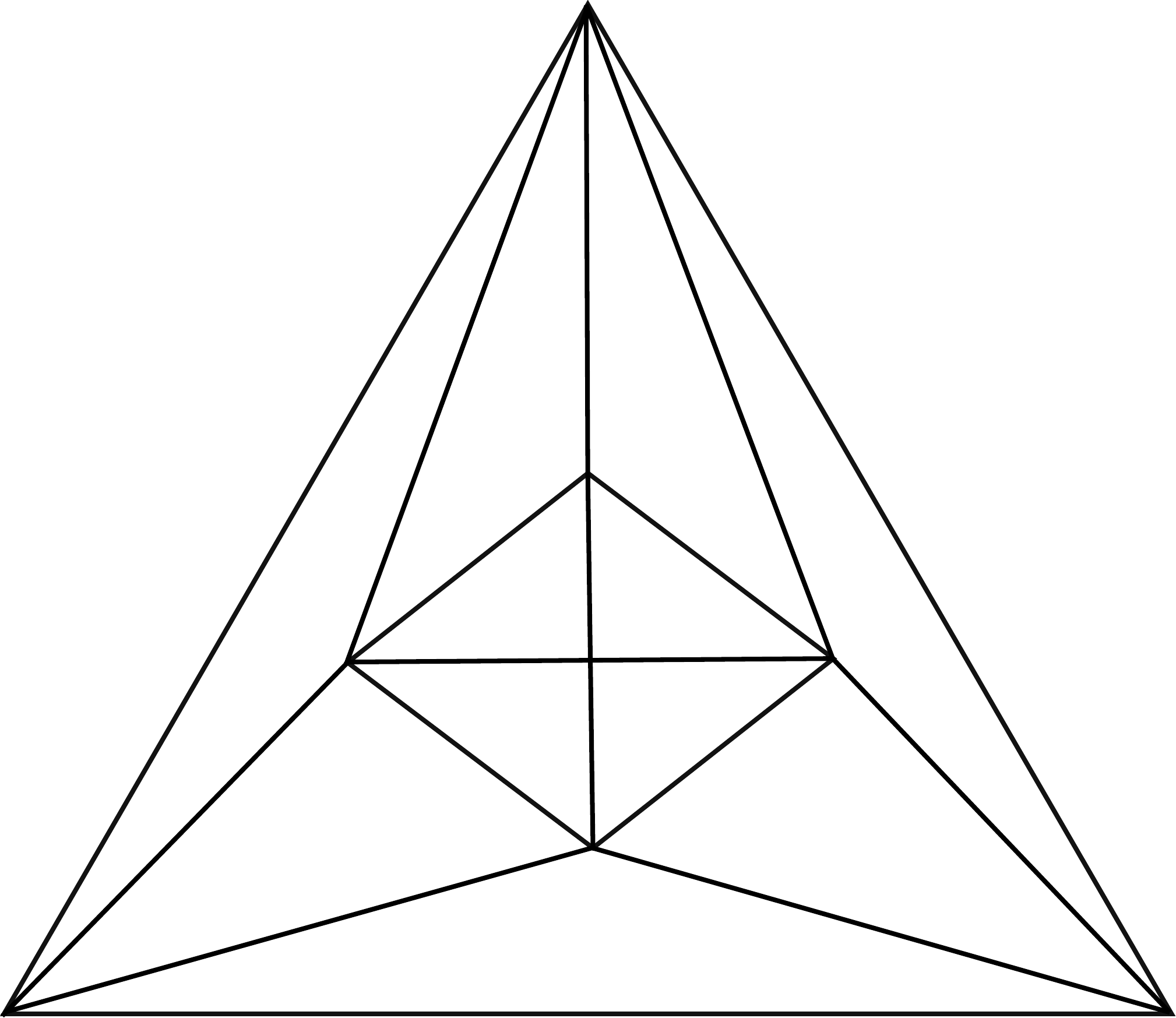}
  \end{center}
\caption{Graph $H_8$.}
\label{S8}
\end{figure}

\begin{figure}
  \begin{center}
    \includegraphics[height=6cm]{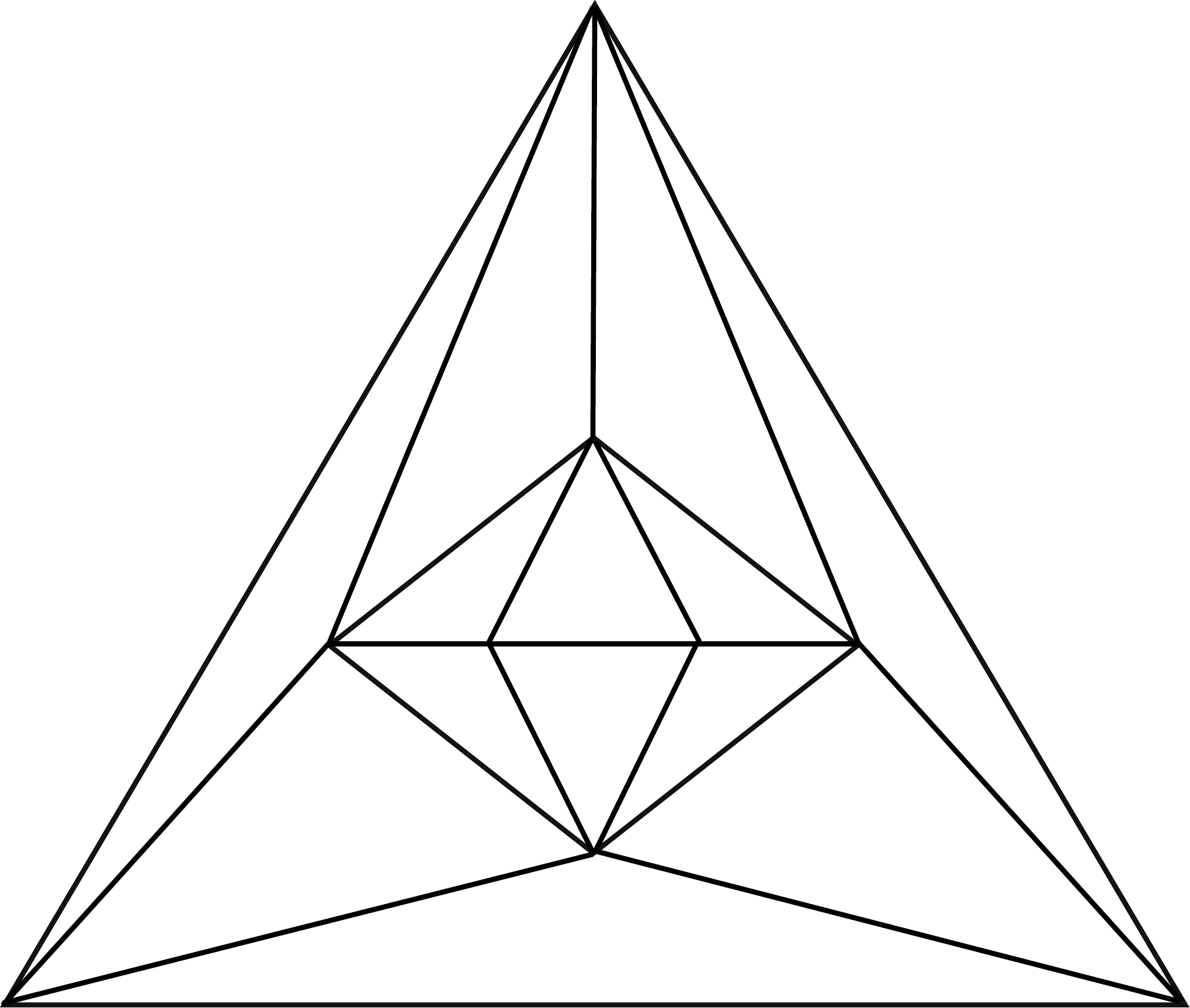}
  \end{center}
\caption{Graph $H_9$.}
\label{S9}
\end{figure}

For comparative purposes, it is valuable to study another recursive family of
planar triangulations with chromatic polynomials of the multi-term form
(\ref{pgform}).  We denote this family as $H_n$, which is well-defined for $n
\ge 8$ and has $n(H_n)=n$. In Fig. \ref{S8} we show the lowest member of this
family, $H_8$. The next higher member, $H_9$, is constructed by adding a vertex
and associated edges in the central ``diamond'', as shown in Fig. \ref{S9}, 
and so forth for higher members.

For this family of planar triangulations we calculate the chromatic polynomial
as 
\beq
P(H_n,q) = \sum_{j=1}^3 c_{H,j}(q)[\lambda_{H,j}(q)]^{n-5} \ , 
\label{psn}
\eeq
where
\beq
c_{H,1}(q) = q(q-3)^3 
\label{cs1}
\eeq
\beq
c_{H,2}(q) = q(q-1)(q^3-9q^2+30q-35)
\label{cs2}
\eeq
and
\beq
c_{H,3}(q) = -q(q-3)(q-5)(q^2-3q+1) 
\label{cs3}
\eeq
Thus, $j_{max}=3$ for this family, as was the case for the $B_n$ family. 
The $\lambda_{H,j}$ are also the same as those for the $B_n$ family, namely 
\beq
\lambda_{H,1}(q) = q-2, \quad \lambda_{H,2}(q)=q-3, \quad 
\lambda_{H,3}(q)=-1 \ . 
\label{lambdasn}
\eeq
$P(H_n,q)$ contains the factor $P(K_4,q)$ and hence has $\chi(H_n)=4$. 

Evaluating $P(H_n,q)$ at $q=\tau+1$, we find 
\beq
P(H_n,\tau+1) = \Big ( \frac{-7+3\sqrt{5}}{2} \ \Big ) \, (\tau-1)^{n-5} + 
\Big ( \frac{5-3\sqrt{5}}{2} \ \Big ) (\tau-2)^{n-5} \ . 
\label{psntp1}
\eeq
Consequently, 
\beq
r(H_n) = \frac{-7+3\sqrt{5}}{2} + \Big ( \frac{5-3\sqrt{5}}{2} \ \Big )
\Big ( \frac{1-\sqrt{5}}{2} \ \Big )^{n-5} \ . 
\label{rsn}
\eeq
Since $|(1-\sqrt{5} \, )/2| < 1$, the second term in
Eq. (\ref{rsn}) vanishes as $n \to \infty$, so 
\beq
\lim_{n \to \infty} r(H_n) = \frac{7-3\sqrt{5}}{2} = 0.145898..
\label{rsninf}
\eeq
and
\beq
a_H = 1 \ . 
\label{as}
\eeq

\section{Nonzero Ground-State Entropy}

In this section we discuss some properties of these families of planar
triangulations relevant for statistical physics.  We recall that the entropy
per vertex is given by $S = k_B \ln W$, where $W$ is the degeneracy per vertex,
related to the total degeneracy of configurations $W_{tot}$ by $W = \lim_{n \to
\infty} (W_{tot})^{1/n}$.  Let us denote the formal limit of a family of graphs
$G$ as $n(G) \to \infty$ by the symbol $\{ G \}$.  The total ground-state entropy
of the Potts antiferromagnet on the graph $G$ is $W_{tot}(G,q ) =
Z_{PAF}(G,q,T=0)=P(G,q)$, so that, formally, $W( \{ G \},q ) = \lim_{n \to
\infty} [P(G,q)]^{1/n}$.  However, owing to the noncommuativity (\ref{wnoncom})
\cite{w}, it is necessary to specify the order of limits in this formal
expression.  For a particular value $q=q_s$, we thus define
\beq
W_{qn}(\{ G \},q_s) = \lim_{q \to q_s} \lim_{n \to \infty} [P(G,q)]^{1/n}
\label{wqn}
\eeq
and
\beq
W_{nq}(\{ G \},q_s) = \lim_{n \to \infty} \lim_{q \to q_s} [P(G,q)]^{1/n} \ . 
\label{wnq}
\eeq
For real $q \ge \chi(G_{pt,m})$, both of these definitions are equivalent, and
in this case we shall write $W_{qn}(\{ G \},q_s) = W_{nq}(\{ G \},q_s) \equiv 
W( \{ G \},q_s)$.  We proceed to discuss these quantities for the families of
graphs studied here. 

For the family $R_m$ with $\chi(R_m)=4$, we have, for $q \ge 4$, 
\beq
W(\{ R \},q) = q-3 \ . 
\label{wr}
\eeq
Since $W(\{ R \},q) > 1$ for $q > 4$, it follows that in this range of $q$ the
Potts antiferromagnet exhibits nonzero ground-state entropy $S_0 = k_B \ln
(q-3)$ on the $m \to \infty$ limit of this family of graphs.

For the family $TC_m$ of cylindrical strips of the triangular lattice, with
$\chi(TC_m)=3$, we have, for real $q \ge 3$, 
\beq
W(\{ TC \},q) = [\lambda_{TC}(q)]^{1/3} \ . 
\label{wtc}
\eeq
This function has the property that $W(\{ TC \},3) = 1$ and $W( \{ TC \},q) >
1$ for real $q > 3$, so that the Potts AF has nonzero ground-state entropy,
i.e., $S_0 = (1/3) k_B \ln [ \lambda_{TC}(q)] > 0$, for $q > 3$ on the
infinite-length limit of this lattice strip. Parenthetically, we note that many
strips of the triangular lattice, such as those with free, cyclic, or M\"obius
boundary conditions with more than one $K_3$, or with toroidal boundary
conditions \cite{baxter87}-\cite{tf} are not planar triangulations, since they
are either not planar or, if planar, they contain at least one face that is not
a triangle, and hence their chromatic polynomials are not subject to the Tutte
bound.

For the family of iterated icosahedra, $I_m$, $\chi(I_m)=4$ and, for $q > 3$,
we calculate
\beq
W_{qn}( \{ I \},q) = [(q-3)f_I(q)]^{1/9}  \ . 
\label{wi}
\eeq
Now $W( \{ I \},q) > 1$ for $q > 3.5133658...$, so that $S_0 = (1/9)k_B \ln
[(q-3)f_I(q)]$ is positive in this interval. Note that $W(\{ I \}, 4) =
(10)^{1/9} = 1.29155..$ 

For the bipyramid family $B_n$, with $\chi(B_n)=3$ for even $n$ and 
$\chi(B_n)=4$ for odd $n$ and for the family $H_n$, with $\chi(H_n)=4$, we 
have $W_{qn}(\{ G \},q) = q-2$, 
and $W_{qn}( \{ G \},q) = W_{nq}( \{ G \},q)$ for $q \ge 4$, where 
$\{ G \} = \{ B \}$ or $\{ H \} $, whence 
\beq
W( \{ G \},q) = q-2 \quad {\rm if} \ q \ge 4 \quad {\rm for} \ \{ G \} = 
\{ B \} \ {\rm or} \ \{ H \} \ . 
\label{wbh}
\eeq
Thus, with the definition $W_{qn}( \{ G \}, q)$, there is nonzero ground-state
entropy $S_0 = k_B \ln(q-2)$ for $q > 3$ and with either definition, 
$W( \{ G \}, q) > 1$ and $S_0 > 0$ for $q \ge 4$ for 
$\{ G \} = \{ B \}$ or $\{ H \} $.

\section{On the Chromatic Zero Nearest to $\tau+1$} 

We have calculated and analyzed the chromatic polynomials of a number of other
planar triangulation graphs. We construct these graphs by starting with a
triangle, $K_3$, inserting a circuit graph with $\ell$ vertices, $C_\ell$,
inside it, connecting the vertices of this circuit graph in an appropriate
manner with the three outer vertices of the original triangle, and then, for
$\ell \ge 4$, tiling the interior of the $C_\ell$ with triangles. 
We have also studied the zeros of the chromatic polynomials of these planar
triangulations, i.e., the chromatic zeros.  Before reporting our new result, we
recall some properties of real chromatic zeros. For a general graph $G$, it is
elementary that there are no negative chromatic zeros. Further, there are no
chromatic zeros in the intervals $(0,1)$ or $(1,32/27]$ \cite{jackson93}.
For a planar triangulation $G_{pt}$, there are no zeros in the interval
$(2,2.546602...)$, where the second number is the unique real zero of
$\lambda_{TC}(q)$ \cite{woodall92}. In 1992, Woodall conjectured that a planar
triangulation has no chromatic zeros in the interval $(q_m,3)$, where $q_m = 
2.6778146..$ is the unique real zero of $q^3-9q^2+30q-35$ 
\cite{woodall92}, but in 2005 he reported that he had found a counterexample to
his conjecture.  We have also found counterexamples, which we shall discuss
below. The four-color theorem states that for a planar graph, $G_p$, $P(G_p,4)
> 0$.  It has been shown that for a particular family of planar graphs given 
in \cite{strip2}, there is an accumulation of real zeros approaching $q=4$ 
from below \cite{royle}. 

It may be recalled that Tutte noticed that many planar triangulations have real
chromatic zeros close to $q=\tau+1$.  We have investigated the following
question: if $G_{pt}$ is a planar triangulation, it it true that the chromatic
zero of $G_{pt}$ nearest to $\tau+1$ is always real?  This is true for almost
all of the families that we have studied; however, we have shown that it is not
true in general. 

\begin{figure}
  \begin{center}
    \includegraphics[height=6cm]{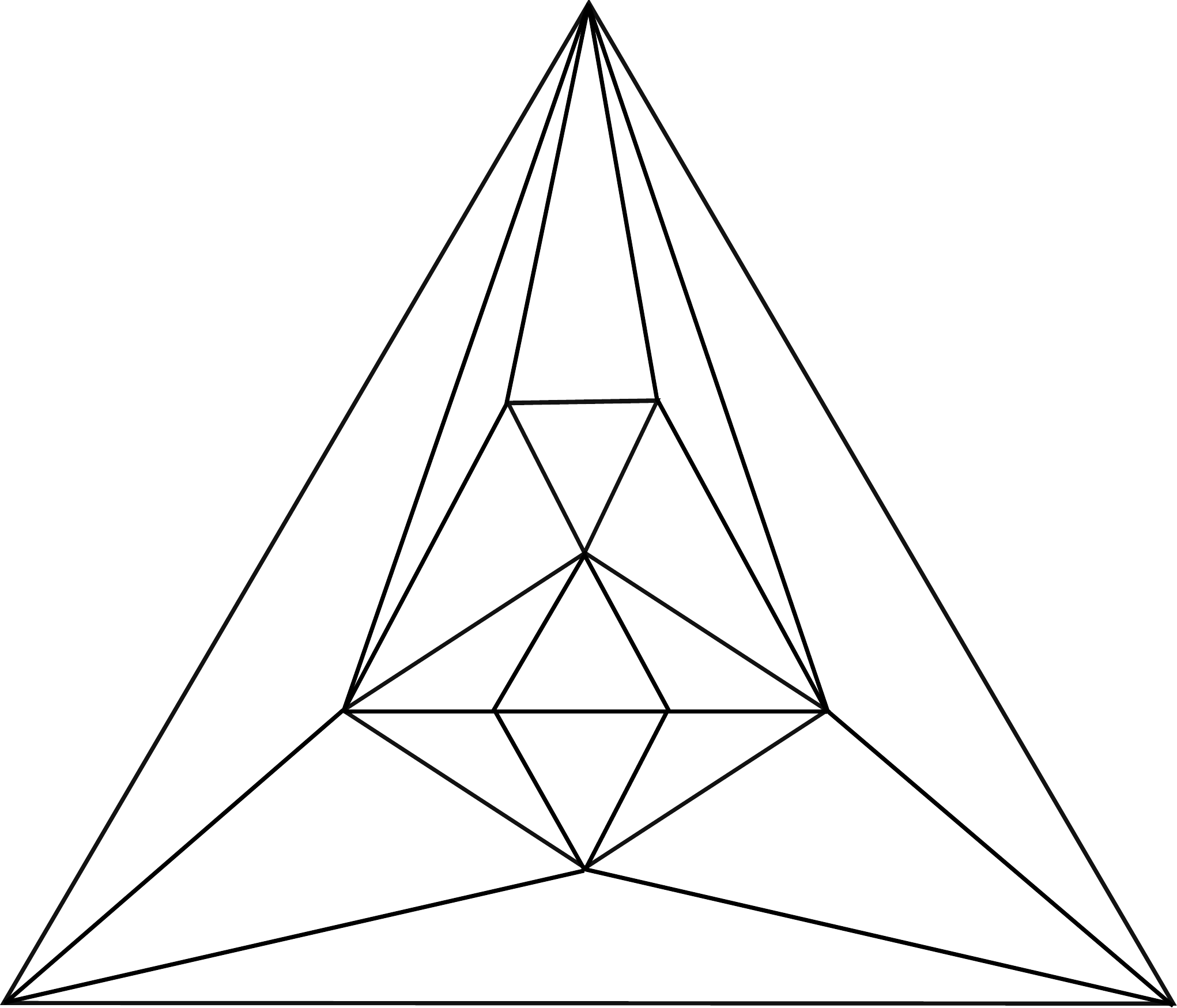}
  \end{center}
\caption{Graph $G_{CM,1}$.}
\label{CM1}
\end{figure}

 We have found a planar triangulation which is, to our knowledge, the first
case for which the zero closest to $\tau+1$ is not real but instead the zeros
closest to $\tau+1$ form a complex-conjugate pair.  We denote this graph as
$G_{CM,m}$ and show a picture of the $m=1$ case in Fig. \ref{CM1}.  The graph
$G_{CM,m}$ has $n(G_{CM,m})=3+8m$.  The chromatic polynomial of this graph has
the form (\ref{pgform1}).  We calculate
\beq
P(G_{CM,m},q) = q(q-1)(q-2) \, [\lambda_{CM}(q)]^m \ , 
\label{pg8b}
\eeq
where
\beqs
f_{CM}(q) & = & q^8-24q^7+259q^6-1641q^5+6674q^4-17818q^3+30418q^2 \cr\cr
          & - & 30250q+13360 \ . 
\label{f8b}
\eeqs
In addition to the zeros at $q=0, \ 1, \ 2$, $P(G_{CM,m},q)$ has eight other
zeros, which form four complex-conjugate pairs. One of these pairs, namely the
zeros at $2.641998 \pm 0.014795i$, lie closest to $\tau+1$, each one being a
distance 0.028163.. from this point.  For reference, 
\beq
r(G_{CM,m}) = \bigg ( \frac{115-51\sqrt{5}}{2} \ \bigg )^m \ . 
\label{rbcm}
\eeq
Hence, 
\beq
a_{G_{CM}} = \bigg ( \frac{115-51\sqrt{5}}{2} \ \bigg )^{1/8} = 0.885185..
\label{agcm}
\eeq

\begin{figure}
  \begin{center}
    \includegraphics[height=6cm]{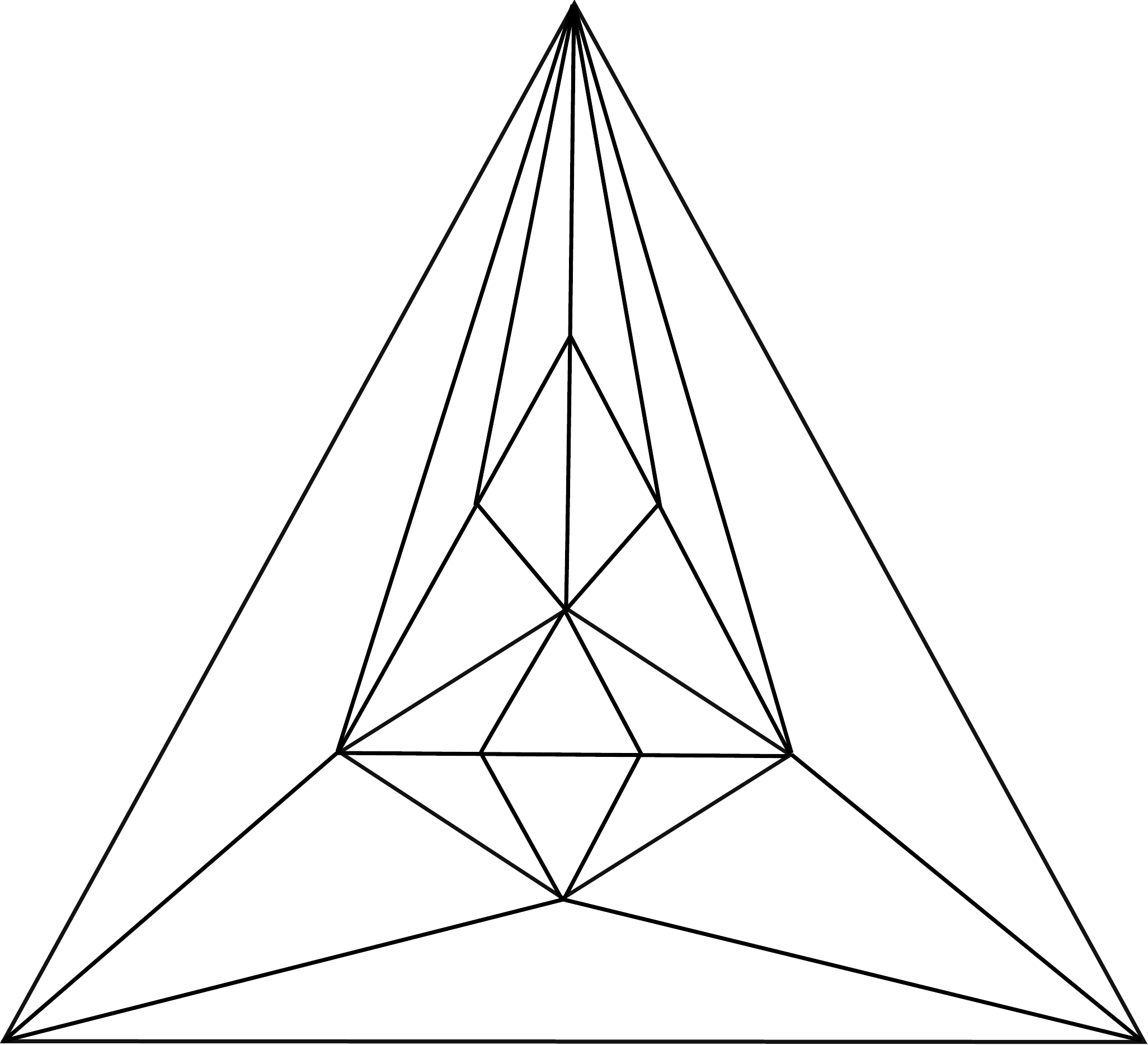}
  \end{center}
\caption{Graph $G_{ce12}$.}
\label{gn12nonwood}
\end{figure}

We have found a number of counterexamples to the 1992 Woodall conjecture that
for a planar triangulation, there are no chromatic zeros in the interval 
$(q_m,3)$.  We show one in Fig. \ref{gn12nonwood}. 
For this graph we calculate the chromatic polynomial
\beqs
P(G_{ce12},q) & = & q(q-1)(q-2)(q-3) \, (q^8-24q^7+260q^6-1658q^5 \cr\cr
              & + & 6800q^4-18337q^3+31668q^2-31915q+14314) \ . 
\label{pgn12nonwood}
\eeqs
Besides its zeros at $q=0, 1, 2, 3$, this chromatic polynomial has two more
real zeros, at $q=2.61461437..$ and $q=2.81889716..$.  The former of these is
close to $\tau+1$, and the latter is notable as lying in the interval 
$(2.677814..., 3)$.  The other zeros of $P(G_{ce12},q)$ form three
complex-conjugate pairs.  We note that 
\beq
r(G_{ce12}) = 101-45\sqrt{5} = 0.376941...
\label{rce12}
\eeq

\section{Conclusions}

In this paper we have studied the ratio of the chromatic polynomial of a planar
triangulation evaluated at $q=\tau+1$ to the Tutte upper bound,
$r(G_{pt})=|P(G_{pt},\tau+1)|/(\tau-1)^{n-5}$, for a variety of $G_{pt}$
graphs. We have constructed and analyzed infinite recursive families of planar
triangulations $G_{pt,m}$ depending on a parameter $m$ linearly related to $n$
and have shown that if $P(G_{pt,m},q)$ only involves a single power of a
polynomial, $[\lambda_{G_{pt}}(q)]^m$, then $r(G_{pt,m})$ approaches zero
exponentially fast as $n \to \infty$. We have analyzed infinite recursive
families for which $P(G_{pt,m},q)$ is a sum of several powers, as in
Eq. (\ref{pgform}) with $j_{max} \ge 2$ and have shown that for these families,
$r(G_{pt,m})$ may approach a finite nonzero constant as $m \to \infty$.  We
have also elucidated the connection between the Tutte upper bound and chromatic
zero(s) near to $\tau+1$. In particular, we have discovered a graph which is,
to our knowledge, the first one for which the zero(s) closest to $\tau+1$ is
not real, but instead is a complex-conjugate pair.  In a final section, we have
discussed interesting connections with ground-state entropy of the Potts
antiferromagnet on the $n \to \infty$ limits of these families of graphs.

\section{Acknowledgments}

This research was partially supported by the grant NSF-PHY-09-69739.

\end{document}